\documentclass[final,5p,times,twocolumn]{elsarticle}

\pdfoutput=1

\usepackage{graphicx}
\usepackage{dcolumn}
\usepackage{bm}
\usepackage{amssymb}
\usepackage{color}
\usepackage{eurosym}
\journal{Applied Surface Science}

\begin{document}
\begin{frontmatter}

\title{Magnetic domain wall pinning in cobalt ferrite microstructures}

\author[i1]{Sandra Ruiz-G\'omez}
\affiliation[i1]{Max-Planck-Institut für Chemische Physik fester Stoffe, 01187 Dresden, Germany}

\author[i2]{Anna Mandziak}
\affiliation[i2]{Solaris National Synchrotron Radiation Centre, Krakow 30-392, Poland}

\author[i3]{Laura Mart\'{\i}n-Garc\'{\i}a}
\affiliation[i3]{Instituto de Química Física ``Rocasolano'', CSIC, Madrid 28006, Spain}

\author[i3]{Jos\'e Emilio Prieto}

\author[i4]{Pilar Prieto}
\affiliation[i4]{Dpto. de Física Aplicada, Universidad Autónoma de Madrid, Madrid 28049, Spain}

\author[i5]{Carmen Munuera}
\affiliation[i5]{Instituto de Ciencia de Materiales de Madrid, CSIC, Madrid 28049, Spain}

\author[i6]{Michael Foerster}
\affiliation[i6]{Alba Synchrotron Light Facility, Cerdanyola del Valles 08290, Spain}

\author[i7]{Adri\'an Quesada}
\affiliation[i7]{Instituto de Cerámica y Vidrio, CSIC, Madrid E-28049, Spain}

\author[i6]{Lucía Aballe}

\author[i3]{Juan de la Figuera}

 \begin{abstract}
  
  A detailed correlative structural, magnetic and chemical analysis of non-stoichiometric cobalt ferrite micrometric crystals was performed by x-ray magnetic circular dichroism combined with photoemission microscopy, low energy electron microscopy, and atomic force microscopy. The vector magnetization at the nanoscale is obtained from magnetic images at different x-ray incidence angles and compared with micromagnetic simulations, revealing the presence of defects which pin the magnetic domain walls. A comparison of different types of defects and the domain walls location suggests that the main source of pinning in these microcrystals are linear structural defects induced in the spinel by the substrate steps underneath the islands.
\end{abstract}

\begin{graphicalabstract}
\includegraphics[width=1\textwidth]{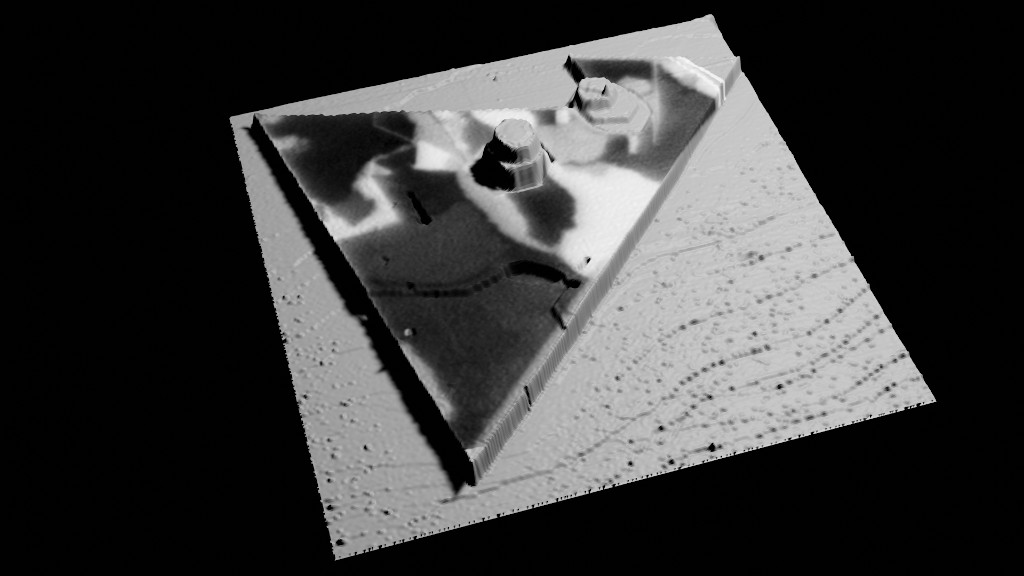}
\end{graphicalabstract}

\begin{highlights}
\item Highly crystalline non-stoichiometric cobalt ferrites have been grown by MBE on Ru(0001).
\item Correlative structural, magnetic and chemical analysis of the crystals have been performed by XMCD-PEEM microscopy, LEEM microscopy, and atomic force microscopy.
\item The source of domain walls pinning has been studied finding a 18\%, 30\% and 45\% of the DWs pinned in chemical defects, AFM features and substrate steps, respectively.
\item The role of substrate steps in pinning magnetic domains is expected to be widespread in magnetic oxide spinels grown on metal substrates.
\end{highlights}


\end{frontmatter}

\renewcommand{\tablename}{Table}

\section{Introduction}

Cobalt ferrite is a ferrimagnetic oxide, which at the stoichiometric composition CoFe$_2$O$_4$ presents a high magnetocrystalline anisotropy compared with other cubic (spinel-based) ferrites\cite{BrabersBook1995}. This property, combined with its high Curie temperature and insulating character has made cobalt ferrite (CFO) popular for spin filtering\cite{CareyAPL2002,ChaplinePRB2006,RamosAPL2007,RigatoPRB2010,TakahashiAPL2010,GuoAPL2016}. CFO thin films, the form required by applications, have been grown by many methods that provide epitaxial layers, among them magnetron sputtering\cite{CareyAPL2002,RigatoPRB2010}, pulsed laser deposition \cite{ChaplinePRB2006,GuoAPL2016} or molecular beam epitaxy \cite{RamosAPL2007}.

Other members of the same family of cubic ferrites with the spinel structure such as magnetite Fe$_3$O$_4$ and NiFe$_2$O$_4$ have also attracted much interest in spintronics. A particular feature of cobalt ferrite is that it can accept a wide range of Fe/Co ratios, which strongly influence its magnetic and electronic properties\cite{takahashiJAP1972,SmithJPCS1978,MoyerPRB2011_XMCD,MoyerPRB2011_XAS}. In Fe-rich compositions\cite{MoyerPRB2011_XMCD,MoyerPRB2011_XAS}, divalent Co cations occupy preferentially octahedral sites while iron cations occupy both octahedral and tetrahedral sites, and depending on the Fe/Co ratio, can present both divalent and trivalent oxidation states.

The detailed magnetic structure of the CFO thin films is highly relevant for their magnetic properties, and are often very different from those of bulk single crystals. This is true even for nominally epitaxial layers of high structural quality. One reason is that films posess a variety of defects that are not present in the bulk material, or at least not in the same densities. For example, many spinel films present high densities of so-called antiphase boundaries (APBs)\cite{EerensteinPhD}. Antiphase boundaries appear when films grow epitaxially on substrates that provide nucleation centers for the spinel phase at distances that are not integer multiples of the spinel unit cell. A classic case is films of spinel oxides grown on MgO, which has a smaller unit cell. In this case, the anion lattice might be continuous throughout the film but antiphase boundaries appear between regions that originate from different nuclei. Many of the unexpected magnetic properties of films of spinel ferrites in general \cite{EerensteinPhD} and cobalt ferrite in particular \cite{MoyerPRB2012} have been attributed to their presence\cite{MarguliesPRL1997}. The effect of APB's on the magnetic domains has been recently observed directly by transmission electron microscopy\cite{KasamaPRB2006}, their structure determined at the atomic level\cite{MckennaNatComm2014} and their magnetic interactions determined through atomistic spin dynamics\cite{MorenoJPc2021}. Since APBs are difficult to remove after growth\cite{EerensteinPRB2003}, deposition methods that avoid their appearance are being sought. One method applied with some success is to employ special substrates\cite{SinghAdvMat2017,SinghJAP2019,SrivastavaIEEEmag2020} with an isostructural spinel unit cell and a very small lattice mismatch. However, this limits the substrates to a few particular oxide spinels, such as CoGa$_2$O$_4$ or MgGa$_2$O$_4$. Another approach is to induce film growth from a single nucleus. Although a continuous film is difficult to grow in such way, we have shown that using high-temperature oxygen assisted molecular beam epitaxy individual islands with sizes of up to tens of micrometers can be obtained, both for magnetite \cite{SantosJPc2009,MontiPRB2012,SandraNano2018}, nickel ferrite \cite{AnnaSciRep2018} or cobalt ferrite \cite{LauraAM2015, SandraJChemPhys2020}. Such microcrystals, each grown from a single nucleus, present magnetic domains in remanence that are orders of magnitude larger that those of films deposited by more standard methods, a finding that was ascribed to the absence of antiphase boundaries.

However, even if the magnetic domains are very large, not all of them can be explained without invoking the presence of defects acting as pinning sites. In this work we consider in detail the possible defects which could cause the observed magnetization domain distribution. We believe our results can be applicable to other oxide spinels thin films.

\section{Experimental Methods}

The growth of the cobalt ferrite crystals and the subsequent x-ray absorption experiments have been performed at the CIRCE experimental station of the Alba synchrotron\cite{CIRCE}. It is equipped with a low-energy electron microscope than can also be used to image the distribution of photoemitted electrons upon x-ray illumination, i.e. as a photoemission microscope. In this mode, it can provide images of the energy-filtered distribution of photoelectrons with a spatial resolution down to 20~nm and an energy resolution down to 0.2~eV. 

The x-ray beam hits the sample at an angle of 16$^\circ$ from the surface plane, while the azimuthal angle of the sample can be changed in order to probe different components of the magnetization. Both for x-ray absorption spectro-microscopy (XAS-PEEM) and for x-ray magnetic circular dichroism microscopy (XMCD-PEEM), we use photoelectrons from the secondary electron background to form the images, i.e., electrons photoemitted from the sample at very low kinetic energies (typically 2~eV).  Dichroic images are obtained from the pixel-by-pixel asymmetry between images acquired with opposite x-ray helicities at the resonant x-ray absorption energies of the magnetic elements\cite{SchneiderPEEM2002}. A single image for a given x-ray beam incidence angle relative to the sample gives only the component of the magnetization along the beam direction. By changing the azimuthal angle between the sample and the x-ray beam, the magnetization can be measured along different directions. All components of the magnetization can be obtained if at least three non-coplanar directions are measured\cite{SandraNano2018}. In our experimental setup the polar angle is fixed to 16$^\circ$ between the x-ray beam and the surface plane. Thus, the setup is more sensitive to the in-plane magnetization components, although out-of-plane magnetization components can also be detected. 

The substrate is a Ru(0001) single crystal cleaned by cycles of annealing in oxygen at 1200~K in 10$^{-6}$~mbar of molecular oxygen, followed by flashing to 1800~K in vacuum. The growth of the mixed cobalt-iron oxides is performed keeping the substrate at high temperature (typically 1100~K) while depositing Co and Fe in a molecular oxygen background pressure of 10$^{-6}$~mbar. Fe and Co are deposited using home-made dosers containing a rod of each material heated by electron bombardment and surrounded by a water cooling jacket. 

The micromagnetic simulations were performed with the MuMax3 software\cite{MuMax3} using a low-end graphic GPU (2Gb GeForce GTX760). We used the bulk materials constants for stoichiometric CFO: saturation magnetization, exchange stiffness and first order magnetocrystalline cubic anisotropy were $M_s= 3\times 10^5$ A m$^{-1}$, $A_{ex}= 2.64\times 10^{-11}$ J m$^{-1}$, and $K_{c1}= 12.5\times 10^4$ J m$^{-3}$, respectively. The cubic anisotropy axis were assigned considering that the islands are (111) terminated, and that the island sides run along the $\langle 110 \rangle$ directions. Each magnetic configuration was relaxed in order to minimize first the energy and then the total torque using a Bogacki-Shampine solver\cite{MuMax3}. The mesh size was 904$\times$765$\times$1 cells and each cell is 8.46 nm $\times$ 8.46 nm $\times$ 3 nm (different values of the height between 3 and 10 nm were employed without any significant difference in the results). The lateral size of the cell was chosen to coincide with the experimental resolution of the images.

\section{Results and discussion}

\begin{figure}[htb]
	\centerline{\includegraphics[width=0.4\textwidth]{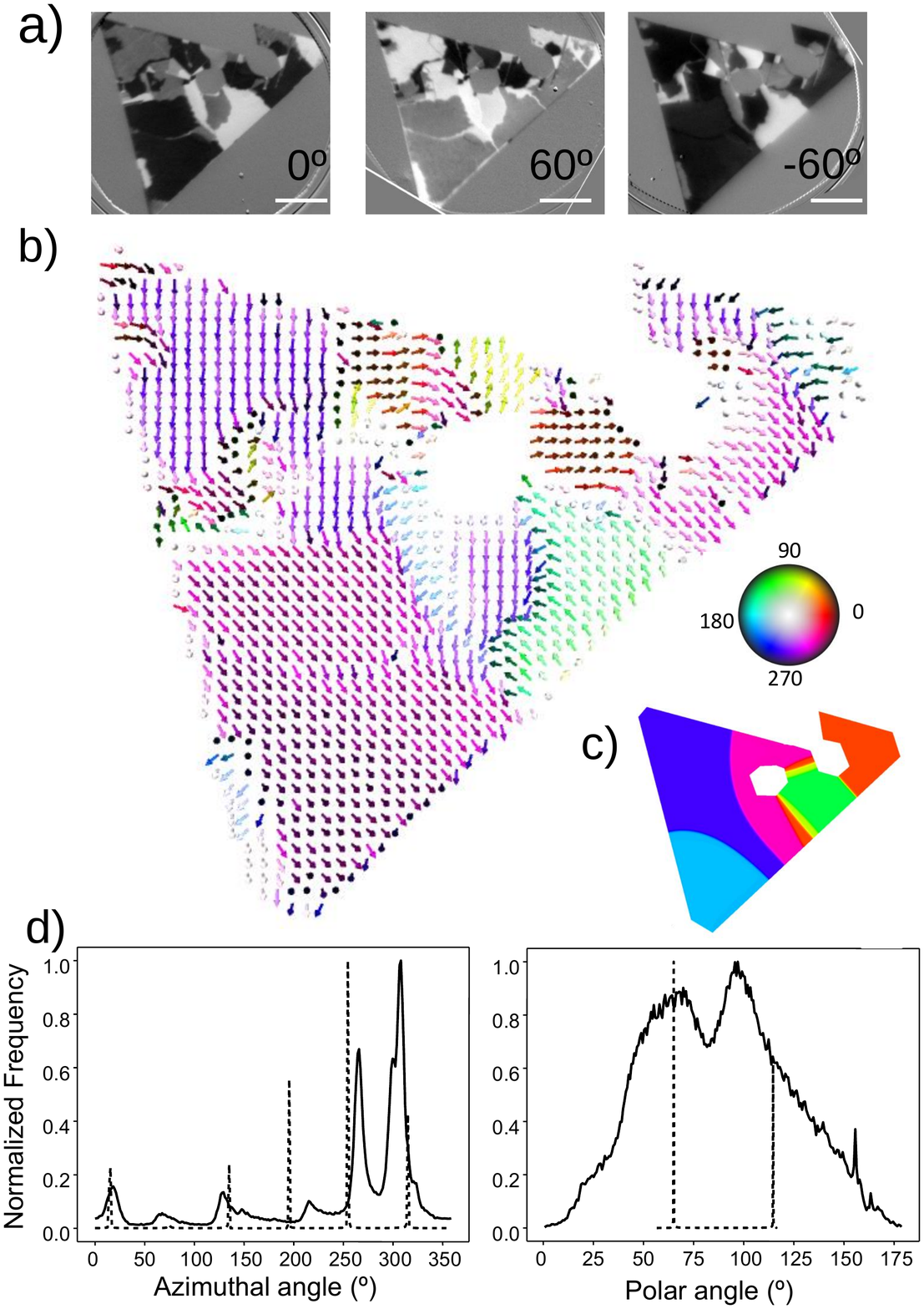}}
	\caption{a) XMCD images of the same island acquired with a photon energy corresponding to the Fe L$_3$ edge, and with azimuthal angles of the x-ray beam relative to the sample of respectively 0, 60 and -60$^\circ$. The scale bar is 2 $\mu m$. b) Arrow representation of the magnetization vector obtained from the previous images. The color of each arrow indicates the in-plane orientation according to the color wheel shown on the right side and the saturation the out-of-plane component, from fully pointing towards the substrate (black) to fully pointing towards the surface (white). c) In-plane micromagnetism relaxed configuration, using a random initial configuration and the material parameters described in the experimental section. The color codes are the same as used in b). The shape and thickness of the islands are taken from the experimentally measured ones. d) Histogram of the magnetization as a function of the in-plane azimuthal angle (where 0$^\circ$ indicates magnetization pointing to the right) and of the polar angle (where 90$^\circ$ indicates in-plane). The pixel size is 8.46 nm both in the simulation and the experimental image. The continuous lines correspond to the experimental data, and the dashed ones to the relaxed micromagnetic simulation.}
	\label{fg:vector}
\end{figure}

The growth of the spinel islands has been described in detail before\cite{LauraAM2015,SandraJChemPhys2020}, and thus here we will just summarize  the most relevant aspects. The growth of mixed iron-rich Co-Fe oxides\cite{LauraPP2016,SandraJChemPhys2020} starts with the nucleation of islands composed of a divalent mixed oxide with the rock salt structure (Fe$_x$Co$_{1-x}$O). Such islands then grow and coalesce to form a complete layer wetting the substrate, typically two atomic layers thick (depending on flux, substrate temperature and oxygen pressure). Upon continuing the deposition, three dimensional islands with the spinel structure nucleate on top of the wetting layer. At high temperature such islands can grow up to hundred nanometers in thickness, and they are typically well separated from each other. We note that due to the high diffusivity at elevated temperatures, the composition of both wetting layer and spinel islands evolves with time, as shown by chemical maps acquired during growth by photoemission microscopy\cite{SandraJChemPhys2020}.

After growth, the sample is brought to room temperature in an oxygen background pressure in order to avoid reduction. The islands composition, determined by XAS-PEEM and their structure, measured by microspot low energy electron diffraction, reveal that they are iron-rich cobalt ferrite (Co$_{0.5}$Fe$_{2.5}$O$_4$\cite{LauraAM2015}). Magnetic mapping is done by means of XMCD-PEEM. Three non-coplanar x-ray incidence angles are measured, as shown in Figure~\ref{fg:vector}a for respectively 0, 60 and -60$^\circ$ azimuthal angle for a representative island of about 10~nm thickness.  In our geometry, the white regions correspond to areas with their local magnetization along the x-ray beam, black ones to areas where the local magnetization is in the opposite direction, and gray regions to areas where there is either no local magnetization or the magnetization is orthogonal to the light direction\cite{SchneiderPEEM2002,LauraAM2015}.  No contrast is detected in the wetting layer, as expected at room temperature. In Figure~\ref{fg:vector}b the obtained magnetization vector is presented. The magnetization is represented by arrows whose color indicates the in-plane orientation and saturation the out-of-plane component. The magnetic domains form a rather intricate landscape with a wide distribution of sizes, and convoluted domain walls.

\begin{figure*}[htb]
	\centerline{\includegraphics[width=1\textwidth]{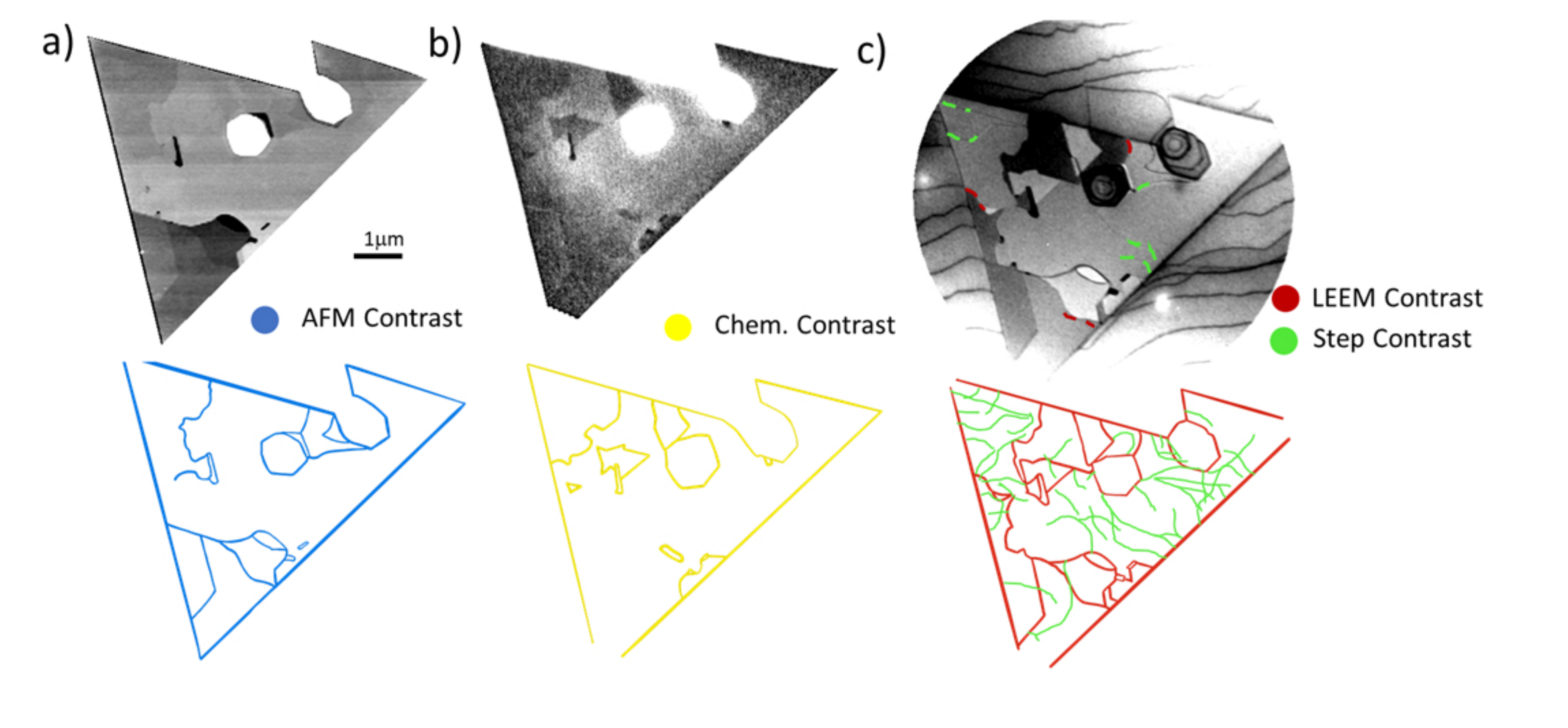}}
	\caption{(a) AFM image acquired ex-situ in the same island. (b) Chemical contrast image at the O K absorption edge. (c) Representative LEEM image acquired in the same island at 23~eV to enhance the contrast of the Ru steps. Bottom row show the draws of the features observed in the images.}
	\label{fg:AFM}
\end{figure*}

The first obvious observation is that the magnetization directions are clearly not following the island edges. In fact, the magnetization direction is often perpendicular to the nearest island edge. This observation rules out shape anisotropy as the main factor determining the magnetization direction, in contrast to what is observed in magnetite islands \cite{SandraNano2018}. The domains walls, like those of magnetite\cite{LauraSciRep2018}, are not chiral. In Figure~\ref{fg:vector}d the histogram of the magnetization as a function of the azimuthal and polar angles through the island is plotted. It is clear that there are several well defined directions in the in-plane orientation of the magnetization, at 19, 68, 131, 216, 265, and 309$^\circ$ which correspond roughly to intervals of 60$^\circ$. From the diffraction patterns and the geometry of the substrate we know that the islands present a (111) surface, and that their edges are oriented along the compact directions of the spinel phase, i.e. the in-plane $\langle110\rangle$ directions. Cobalt ferrite in the bulk form presents cubic anisotropy\cite{BrabersBook1995}, with the easy axes along the $\langle100\rangle$ directions. There is no easy axis within the (111) plane, so we consider instead the projection of the bulk easy axes direction on the (111) plane, which are the in-plane $\langle112\rangle$ directions. Since the composition of the islands lies between that of stoichiometric CoFe$_2$O$_4$ and that of magnetite,  we must also consider the easy axes of pure magnetite at room temperature, which are the $\langle111\rangle$. However, the projection of the $\langle111\rangle$ directions on the (111) plane also corresponds to the in-plane $\langle112\rangle$ directions, so the same orientation is expected. In fact, the experimental magnetization directions roughly correspond to the $\langle112\rangle$ ones, i.e. the bisectrices of the epitaxial triangular islands. This, together with the significant out-of-plane component present, leads to the conclusion that magnetocrystalline anisotropy is the main responsible for the experimentally observed magnetization directions. 

In order to better understand the origin of the magnetic domain distribution observed in our cobalt ferrite islands, we performed micromagnetic simulations using the MuMax3 code\cite{MuMax3} on islands with the experimental geometry. Relaxing from a random configuration gives rise to a magnetization distribution quite different from the experimental one that can be seen in Figure~\ref{fg:vector}c. A comparison of experimental and calculated magnetization directions is shown in Figure~\ref{fg:vector}d in the form of continuous histograms and dashed lines, respectively. While the domain distribution cannot be reproduced by the simulation, the orientation of the domains is in reasonable agreement with the micromagnetic simulations, as expected if magnetocristalline anisotropy is driving the orientation of the magnetization within each domain. However, it is clear that assuming a structurally perfect island, as done in the micromagnetic simulations, does not correctly reproduce the experimental domain distribution.

\begin{figure*}[htb]
	\centerline{\includegraphics[width=0.4\textwidth]{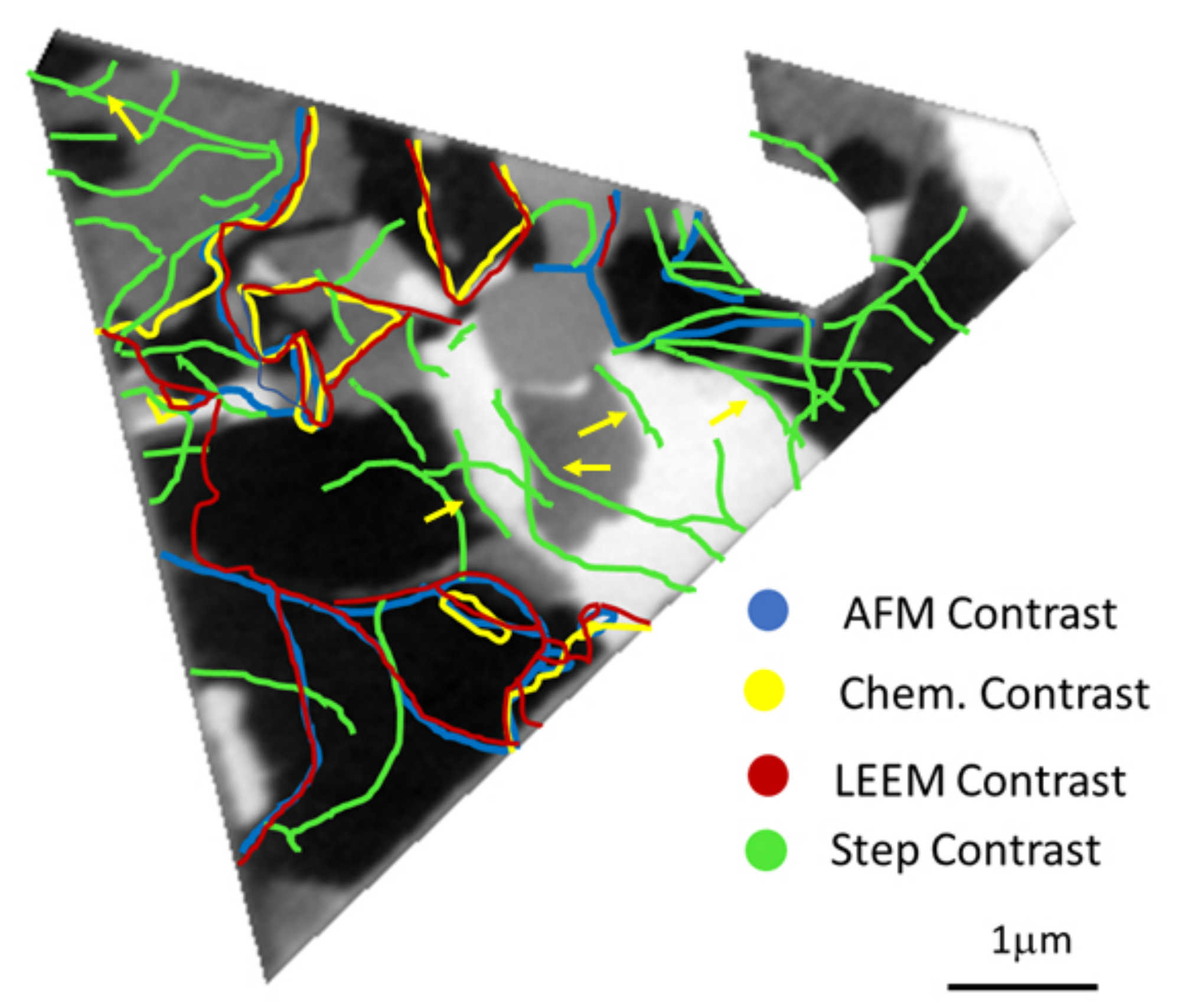}}
	\caption{ Comparison of magnetic domains (image), topographic AFM contrast (blue lines), chemical contrast (yellow lines), LEEM contrast (red lines), and  substrate steps (green lines). Yellow arrows mark regions in which domain walls are pinned in substrate steps.}
	\label{fg:contrast}
\end{figure*}
A straightforward explanation for the observed domain distribution  (see Figure~\ref{fg:AFM}a) could be that the magnetic domain walls get pinned on linear defects. The spinel islands might present different types of linear defects such as (i) steps at the island surface, (ii) boundaries between regions with different composition, (iii) steps at the substrate-island interface, and boundaries (iv) of stacking faults, (v) between antiphase domains, or (vi) between twin domains. We start by locating each of them in order to discriminate which ones are responsible for the pining of domain walls.

Steps at the island surface were unambiguosly located by ex-situ AFM on the very same areas observed by PEEM (see Figure~\ref{fg:AFM}a). The island is resting on a region with wavy substrate steps running mostly along the x-axis. The location of substrate steps can be determined through the Co$_x$Fe$_{1-x}$O wetting bilayer\cite{IreneJPc2013}. The island has nucleated close to a small hexagonal protrusion, which is covered with Co$_x$Fe$_{1-x}$O but not by the spinel island. The top of the spinel island is remarkably flat, with only some steps of around 0.4 nm height at the north-west corner. This is the expected height for atomic steps of a spinel phase along the $\langle111\rangle$ direction. In addition, the area around the south corner is about 1~nm lower than the rest of the island, and presents additional steps. Otherwise, the island has an atomically flat top surface, and therefore a cross-sectional wedge shape, with 12~nm thickness at the south part, and 8~nm near the northern side. 

To correlate possible changes in composition with the location of the domain walls, images with chemical contrast can be obtained by averaging the two XAS-PEEM images acquired with opposite circular polarization. We note that the intensity of the XAS signal has been integrated beyond the white line, in order to avoid any leakage from the magnetic signal. No chemical contrast whatsoever is observed at either the images at the Fe or Co absorption edges (not show). Nevertheless, at the post-O K edge (shown in Figure~\ref{fg:AFM}b), some regions with different contrast are visible. The change in contrast is 5\%. A possible source for some of the observed different gray levels is that they arise from differences in the total height of the island, as some coincide with surface or substrate steps. 

In low-energy electron microscopy mode, an elastically scattered electron beam of a selected energy is used to form an image of the surface (see Figure~\ref{fg:AFM}c). Varying the electron energy and the focusing conditions can provide contrast due to thickness or to several types of buried defects. A classic example is the observation of interface steps. The substrate steps underneath Ag islands on Si(111)\cite{TrompPRL1993} were observed using slightly out-of-focus conditions. The observation was explained in terms of the long range strain fields associated with the interface steps, visible in islands several nanometers thick. In the uniform gray areas of the island (see Figure~\ref{fg:AFM}c), faint lines can be distinguished to follow the paths of the substrate steps coming from outside the island. It is reasonable to presume that they continue below the island giving rise to the observed lines in the LEEM image of the island.

Another type of defects that can be detected by LEEM imaging through the islands are stacking faults. For example, it has been shown that regions with different stacking sequence in Co islands on Ru(0001)\cite{FaridNJP2007} present different electron reflectivities at a given energy. There are several regions on the LEEM image of the island that show different gray level contrast at several energies (see Figure~\ref{fg:AFM}c). While a distinction between different stacking fault types cannot be done without additional information, we believe it is reasonable to interpret the borders between such regions as extended defects, which in a spinel structure can correspond to several types of stacking faults that can be considered to 
consist of combinations of three basic types, called I, II and III in Ref.~[\cite{HornstraJPCS1960}].

Figure~\ref{fg:contrast} shows the comparison of magnetic domains (image) together with the topographic AFM contrast (blue lines), chemical contrast (yellow lines), LEEM reflectivity contrast (red lines), and  substrate steps (green lines). Evaluating the \% of domain wall length that correspond to a change in contrast in the different images we observe the following: 18\% of the total DW length correspond to boundaries in the chemical contrast image (yellow lines), 30\% are located in features observed in the AFM image (blue lines) and coincide with boundaries observed in the LEEM image (red lines) and 45\% are located over substrate steps (see yellow arrows in Figure~\ref{fg:contrast}), although there are many more steps than magnetic domain walls.

We thus find that most magnetic domain walls are located over substrate steps. We analyze now the reasons why substrate steps might strongly pin magnetic domain walls. Again we remark that we lack a detailed atomic information of the matching of the spinel island to the steps of the Ru substrate. However, we can discuss the general structure of the island and the substrate. The Ru(0001) substrate 
has an hcp stacking sequence with an interplanar distance of 0.214 nm, which is the height of the monoatomic steps at the surface. 
Spinel islands along a $\langle111\rangle$ direction can be considered to have an fcc stacking sequence of units, each one composed of a cation layer and a close-packed oxygen layer and with a height close to 0.242~nm. There are two different types of such units which alternate along the vertical direction and differ in the composition and structure of the cation layer, which can be either a mixed tetragonal-octahedral cation layer or a kagomé octahedral cation layer\cite{HornstraJPCS1960}.

Thus there is a 13\% difference in height between a Ru layer and a spinel unit. On the other hand,
two such units, one of each type, are required to form what can be considered the elementary constituent of the spinel structure, so that
the presence of a substrate step underneath an epitaxial island can give rise to adjacent non-equivalent spinel units building either an extended defect along a line parallel to the step
or stacking faults of different types\cite{HornstraJPCS1960}, including possibly twins. We note that for most substrate steps underneath the spinel island we do not find large extended defects within our experimental resolution. However, we cannot rule out the presence of narrow ribbons of stacking faults around the substrate steps (expected to be a few nm wide\cite{FigueraPRB2001}).

The effect of a monoatomic step underneath an epitaxial island can be quantified by the strain field arising from the 13\% difference in height causing magnetostriction in the spinel phase. This has been measured for magnetite, where a typical microcoercivity in the range of mT for bulk dislocations has been theoretically predicted\cite{XuJGR1989} and experimentally observed\cite{LindquistJGR2015}. A detailed prediction of the effect of the stacking faults induced by the substrate steps requires atomistic spin calculations as well as detailed atomistic models, but we can expect effects comparable to those caused by antiphase boundaries on the magnetic properties of spinels. 
Antiphase boundaries in epitaxial (100)-oriented spinel (magnetite) films produce antiferromagnetic 180$^{\rm o}$ superexchange interactions between octahedral cations that oppose their natural ferromagnetic coupling and therefore locally weaken the effective exchange stiffness, thus favouring the pinning of the domain wall at these defects\cite{VoogtPRB1998}. Calculations have been performed for antiphase boundaries in particular geometries\cite{MorenoJPc2021} and it has been reported that they can produce strong pinning of the domain walls.
In some of the possible types of stacking faults in (111)-oriented spinel films, such 180$^{\rm o}$ superexchange interactions are also present, as in the S-II case\cite{HornstraJPCS1960}, where oxygen atoms are normally stacked and the fault affects only the cation layers. Furthermore, they can also appear at the boundaries between differently stacked regions, particularly if one of these is also of type S-II.  
Thus, we suggest that the main cause for the pinning of domain walls in our epitaxial cobalt ferrite islands on Ru(0001) is ultimately the coupling mismatch imposed by the substrate steps.

\section{Conclusions}
By experimentally locating the positions of magnetic domain walls in highly perfect non-stoichiometric cobalt ferrite islands, correlating with the positions of chemical and structural defects and comparing with micromagnetic simulations, we find that the magnetic domain walls are pinned at linear structural defects. 45\% of the domain walls are found on top of interface steps, which we thus conclude are the defects ultimately responsible for the pinning of domain walls.  

We suggest that interface steps play a similar role in (111) oriented spinels grown on hexagonal metal substrates such as Pt(111) and Ru(0001) to antiphase boundaries found on spinel crystals grown on MgO and other square-symmetric substrates. However, contrary to the case of antiphase boundaries which stem from the coalescence of islands nucleated on the same terrace, the defects responsible for the pinning of the domain walls are the substrate steps. It should be thus possible to obtain structurally perfect ferrimagnetic crystals with domain sizes limited only by the substrate terrace size. 

\section{Acknowledgment}

This work is supported by the Grants RTI2018-095303-B-C51,-A-52, and -B-C53  funded by MCIN/AEI/10.13039/501100011033 and by “ERDF A way of making Europe”, and by the Grant  S2018-NMT-4321 funded by the Comunidad de Madrid and by “ERDF A way of making Europe”. These experiments were performed at the CIRCE beamline of the ALBA Synchrotron Light Facility.

\bibliographystyle{elsarticle-num-names} 
\bibliography{cfo}

\end{document}